\documentstyle[12pt]{article}

\newcommand{\np}[3]{{\sl Nucl. Phys.} {\bf #1} (19#2)~#3}
\newcommand{\pl}[3]{{\sl Phys. Lett.} {\bf #1} (19#2) #3}
\newcommand{\pr}[3]{{\sl Phys. Rev.} {\bf #1} (19#2) #3}
\newcommand{\ijm}[3]{{\sl Int. J. Mod. Phys.} {\bf #1} (19#2) #3}
\newcommand{\hep}[1]{{\sl hep--ph/}{#1}}

\begin{document}

\begin{center}
\vspace{\baselineskip}%

{\Large {\bf Restoring gauge invariance in gauge-boson production processes}}\\

\vspace{\baselineskip}%

W. Beenakker~\footnote{Research supported by a fellowship of the Royal Dutch 
                       Academy of Arts and Sciences}\\

\vspace{\baselineskip}%

{\it Instituut--Lorentz, University of Leiden, The Netherlands}

\vspace{2\baselineskip}%
{\bf Abstract}
\end{center}
A survey is given of the various gauge-invariance-related aspects that play a 
role when dealing with unstable gauge bosons.  

\vspace{\baselineskip}%

\section{\large {\bf Unstable gauge bosons: lowest order}}

The physics goals of LEP2 and the next linear collider (NLC) cover a large 
variety of topics, e.g.~the determination of the W-boson mass, establishing the
Yang--Mills character of the triple gauge-boson couplings, the search 
for the Higgs boson, the search for supersymmetric particles, a detailed 
study of the symmetry-breaking mechanism, etc. Most of these studies require a 
careful investigation of processes with photons and/or fermions in the initial
and final state. 

If complete sets of graphs contributing to such a process are taken into
account, the associated matrix elements are in principle gauge-invariant.
However, the massive gauge bosons that appear as intermediate particles can 
give rise to poles $1/(k^2-M^2)$ if they are treated as stable particles. 
This can be cured by introducing the finite decay width in one way or another, 
while at the same time preserving gauge independence and, through a proper 
high-energy behavior, unitarity. In field theory, such widths arise naturally 
from the imaginary parts of higher-order diagrams describing the gauge-boson 
self-energies, resummed to all orders. This procedure has been used with great 
success in the past: indeed, the $Z$ resonance can be described to very high 
numerical accuracy. However, in doing a Dyson summation of self-energy graphs, 
we are singling out only a very limited subset of all the possible higher-order
diagrams. It is therefore not surprising that one often ends up with a result 
that retains some gauge dependence. 

Till recently two approaches for dealing with unstable gauge bosons were 
popular in the construction of lowest-order Monte Carlo generators. The first 
one involves the systematic replacement $1/(k^2-M^2) \to 1/(k^2-M^2+iM\Gamma)$,
also for $k^2<0$. Here $\Gamma$ denotes the physical width of the gauge boson
with mass M and momentum $k$. This scheme is called the `fixed-width scheme'.
As in general the resonant diagrams are not gauge-invariant by themselves, 
this substitution will destroy gauge invariance. Moreover, it has no 
physical motivation, since in perturbation theory the propagator for 
space-like momenta does not develop an imaginary part. Consequently, 
unitarity is violated in this scheme. To improve on the latter another 
approach can be adopted, involving the use of a running width $iM\Gamma(k^2)$ 
instead of the constant one $iM\Gamma$ (`running-width scheme'). This,
however, still cannot cure the problem with gauge invariance. 

At this point one might ask oneself the 
legitimate question whether the gauge-breaking terms are numerically relevant 
or not. After all, the gauge breaking is caused by the finite decay width and
is, as such, in principle suppressed by powers of $\Gamma/M$. From LEP1 we 
know that gauge breaking can be negligible for all practical purposes. 
However, the presence of small scales can amplify the gauge-breaking terms.
This is for instance the case for almost collinear space-like photons or
longitudinal gauge bosons at high energies, involving scales of 
${\cal O}(p_{_B}^2/E_{_B}^2)$ (with $p_{_B}$ the 
momentum of the involved gauge boson). In these situations the external 
current coupled to the photon or to the longitudinal gauge boson becomes 
approximately proportional to $p_{_B}$. In other words, in these 
regimes sensible theoretical predictions are only possible if the amplitudes 
with external currents replaced by the corresponding gauge-boson momenta 
fulfill appropriate Ward identities.

In order to substantiate these statements, a truly gauge-invariant scheme is
needed. It should be stressed, however, that any such scheme is arbitrary to a 
greater or lesser extent: since the Dyson summation must necessarily be taken 
to all orders of perturbation theory, and we are not able to compute the 
complete set of {\it all} Feynman diagrams to {\it all} orders, the various 
schemes differ even if they lead to formally gauge-invariant results. Bearing 
this in mind, we need some physical motivation for choosing a particular 
scheme. In this context two options can be mentioned, which fulfill the 
criteria of gauge invariance and physical motivation. 

The first option is the so-called `pole scheme'~\cite{Veltman,Stuart,Aeppli}. 
In this scheme one decomposes the complete amplitude according to the pole 
structure by expanding around the poles 
(e.g.~$f(k^2)/(k^2-M^2) = f(M^2)/(k^2-M^2) + \mathrm{finite~terms}$). As the 
physically observable residues of the poles are gauge-invariant, gauge
invariance is not broken if the finite width is taken into account in the pole 
terms $\propto 1/(k^2-M^2)$. It should be noted, however, that there exists 
some controversy in the literature \cite{Aeppli,Stuart2} about the `correct' 
procedure for doing this and about the range of validity of the pole scheme, 
especially in the vicinity of thresholds. 

The second option is based on
the philosophy of trying to determine and include the minimal set of Feynman 
diagrams that is necessary for compensating the gauge violation caused by the 
self-energy graphs. This is obviously the theoretically most satisfying 
solution, but it may cause an increase in the complexity of the matrix 
elements and a consequent slowing down of the numerical calculations. For the 
gauge bosons we are guided by the observation that the lowest-order decay 
widths are exclusively given by the imaginary parts of the fermion loops in 
the one-loop self-energies. It is therefore natural to perform a Dyson 
summation of these fermionic one-loop self-energies and to include the other 
possible one-particle-irreducible fermionic one-loop corrections 
(``fermion-loop scheme'')~\cite{BHF1}. For the LEP2 process $e^+e^- \to 4f$ 
this amounts to adding the fermionic triple gauge-boson vertex corrections. 
The complete set of fermionic contributions forms a gauge-independent subset 
and obeys all Ward identities exactly, even with resummed 
propagators~\cite{BHF2}. 
As mentioned above, the validity of the Ward identities guarantees a proper 
behavior of the cross-sections in the presence of collinear photons and at 
high energies in the presence of longitudinal gauge-boson modes. On top of 
that, within the fermion-loop scheme the appropriately renormalized matrix 
elements for the generic LEP2 process $e^+e^- \to 4f$ can be formulated in 
terms of effective Born matrix elements, using the familiar language of 
running couplings~\cite{BHF2}.

A numerical comparison of the various schemes~\cite{BHF1,BHF2} confirms the 
importance of not violating the Ward identities. For the LEP2 process 
$e^+e^- \to e^-\bar{\nu}_e\,u\bar{d}$, a process that is particularly 
important for studying triple gauge-boson couplings, the impact of violating 
the $U(1)$ electromagnetic
gauge invariance was demonstrated~\cite{BHF1}. Of the above-mentioned schemes 
only the running-width scheme violates $U(1)$ gauge invariance. The associated 
gauge-breaking terms are enhanced in a disastrous way by a factor of 
${\cal O}(s/m_e^2)$, in view of the fact that the electron may emit a virtual
(space-like) photon with $p_\gamma^2$ as small as $m_e^2$. A similar 
observation can be made at high energies when some of the 
intermediate gauge bosons become effectively longitudinal. There too the 
running-width scheme renders completely unreliable results~\cite{BHF2}. In 
processes involving more intermediate gauge bosons, e.g.~$e^+e^- \to 6f$,
also the fixed-width scheme breaks down at high energies as a result of
breaking $SU(2)$ gauge invariance.

\section{\large{\bf Unstable gauge bosons: radiative corrections}}

By employing the fermion-loop scheme all one-particle-irreducible fer\-mionic 
one-loop corrections can be embedded in the tree-level matrix elements. This 
results in running couplings, propagator functions, vertex functions, etc. 
However, there is still the question about the bosonic corrections. A large 
part of these bosonic corrections, as e.g.~the leading QED corrections, 
factorize and can be treated by means of a convolution, using the 
fermion-loop-improved cross-sections in the integration kernels. This allows 
the inclusion of higher-order QED corrections and soft-photon 
exponentiation. In this way various important effects can be covered. 
Nevertheless, the remaining bosonic corrections
can be large, especially at high energies~\cite{LEP2WW,WWreview}.

In order to include these corrections one might attempt to extend the 
fermion-loop scheme. In the context of the background-field method a Dyson 
summation of bosonic self-energies can be performed without violating
the Ward identities~\cite{BFM}. However, the resulting matrix elements depend
on the quantum gauge parameter at the loop level that is not completely 
taken into account. As mentioned before, the perturbation series has to be 
truncated; in that sense the dependence on the quantum gauge parameter could be
viewed as a parametrization of the associated ambiguity. 

As a more appealing strategy one might adopt a hybrid scheme, adding the 
remaining bosonic loop corrections by means of the pole scheme. This is 
gauge-invariant and contains the well-known bosonic corrections for the 
production of on-shell gauge bosons (in particular W-boson pairs). Moreover, 
if the quality of the pole scheme were to degrade in certain regions of 
phase-space, the associated error is reduced by factors of $\alpha/\pi$. 
It should be noted that the application of the pole scheme to photonic 
corrections requires some special care, because in that case terms 
proportional to $\log(k^2-M^2)/(k^2-M^2)$ complicate the pole
expansion~\cite{Aeppli,WWreview}.


\begin{thebibliography}{99}

\bibitem{Veltman}
M.~Veltman, {\sl Physica} {\bf 29} (1963) 186.

\bibitem{Stuart}
R.G.~Stuart, \pl{B262}{91}{113}.

\bibitem{Aeppli}
A.~Aeppli {\it et al.}, \np{B428}{94}{126}.

\bibitem{Stuart2}
R.G.~Stuart, Univ. of Michigan preprint UM-TH-96-05, \hep{9603351}.

\bibitem{BHF1}
E.N.~Argyres {\it et al.}, \pl{B358}{95}{339}.

\bibitem{BHF2}
W.~Beenakker {\it et al.}, \hep{9612260}.

\bibitem{LEP2WW}
W.~Beenakker {\it et al.}, in {\sl Physics at LEP2},
eds.\ G.~Altarelli, T.~Sj\"ostrand and F.~Zwirner,
(CERN 96-01, Gen\`eve, 1996) Vol.~1, p.~79,
\hep{9602351}.

\bibitem{WWreview}
W.~Beenakker and A.~Denner, \ijm{A9}{94}{4837}.

\bibitem{BFM}
A.~Denner and S.~Dittmaier, \pr{D54}{96}{4499}.

\end{thebibliography}
\end{document}